\documentclass[twocolumn]{aastex631}
\usepackage{listings}
\shorttitle{Efficient Sparse Matrix Astrochemistry}
\shortauthors{Motoyama et al.}

\begin{document}

\title{An Efficient Algorithm for Astrochemical Systems Using Stoichiometry Matrices}

\author[0000-0002-5104-7031]{Kazutaka Motoyama}
\affiliation{Shiga University of Medical Science, Seta, Tsukinowa-cho, Otsu, Shiga 520-2192, Japan}
\affiliation{Institute of Astronomy and Astrophysics, Academia Sinica,  Taipei 10617, Taiwan}
%\affiliation{National Institute of Informatics, 2-1-2 Hitotsubashi, Chiyoda-ku, Tokyo 101-8430, Japan}
%\affiliation{Theoretical Institute for Advanced Research in Astrophysics, Academia Sinica, Taipei 10617, Taiwan}

\author[0000-0001-5557-5387]{Ruben Krasnopolsky}
\affiliation{Institute of Astronomy and Astrophysics, Academia Sinica,  Taipei 10617, Taiwan}
%\affiliation{Theoretical Institute for Advanced Research in Astrophysics, Academia Sinica, Taipei 10617, Taiwan}

\author[0000-0001-8385-9838]{Hsien Shang}
\affiliation{Institute of Astronomy and Astrophysics, Academia Sinica,  Taipei 10617, Taiwan}
%\affiliation{Theoretical Institute for Advanced Research in Astrophysics, Academia Sinica, Taipei 10617, Taiwan}

\author[0000-0003-3339-3341]{Kento Aida}
\affiliation{National Institute of Informatics, 2-1-2 Hitotsubashi, Chiyoda-ku, Tokyo 101-8430, Japan}

\author[0009-0007-9519-4563]{Eisaku Sakane}
\affiliation{National Institute of Informatics, 2-1-2 Hitotsubashi, Chiyoda-ku, Tokyo 101-8430, Japan}

\correspondingauthor{Kazutaka Motoyama and Hsien Shang}
\email{motoyama@belle.shiga-med.ac.jp, shang@asiaa.sinica.edu.tw}

\begin{abstract}
Astrochemical simulations are a powerful tool for revealing chemical evolution in the interstellar medium.
Astrochemical calculations require efficient processing of large matrices for the chemical networks. The large chemical reaction networks often present bottlenecks for computation because of time derivatives of chemical abundances. We propose an efficient algorithm using a stoichiometry matrix approach in which this time-consuming part is expressed as a loop, unlike the algorithm used in previous studies. Since stoichiometry matrices are sparse in general, the performances of simulations with our algorithm depend on which sparse-matrix storage format is used. We conducted a performance comparison experiment using the common storage formats, including the coordinate (COO) format, the compressed column storage (CCS) format, the compressed row storage (CRS) format, and the Sliced ELLPACK (SELL) format. Experimental results showed that the simulations with the CRS format are the most suitable for astrochemical simulations and about three times faster than those with the algorithm used in previous studies. In addition, our algorithm significantly reduces not only the computation time but also the compilation time.
We also explore the beneficial effects of parallelization and sparse-matrix reordering in these algorithms.
\end{abstract}

% #################################################
%  Introduction
% #################################################

\section{Introduction} \label{sec:intro}

Interstellar molecules are observed in a wide range of environments in interstellar space, and their emissions are used as probes of the physical conditions of astronomical objects. 
%Astrochemistry plays an essential role in revealing the evolution of abundances of molecules in interstellar space. 
More than 240 different molecules have been detected in the interstellar and circumstellar medium \citep{2022ApJS..259...30M}. In particular, a wide variety of molecules, including complex organic molecules, are observed in star-forming regions \citep[see][and references therein]{2020ARA&A..58..727J}. During the gravitational collapse of the molecular core, chemical reactions in the gas phase and on the dust grain surface produce complex molecules from simpler molecules. Recent high-resolution observations with ALMA have revealed chemical structures of protoplanetary disks \citep{2021ApJS..257....1O}. Molecules in protoplanetary disks provide material for the atmospheres of newly forming planets and are related to the origin of life.

% Description of PDR chemistry and UMIST database of the ISM.
Astrochemists model the chemical evolution through reaction networks to understand the formation and destruction of molecules observed in assorted sources in the ISM. In these models, the chemistry is described by processes of various types: ion–neutral and neutral–neutral reactions, dissociation and ionization by UV photons and cosmic-ray particles, dissociative recombination, etc. The models compute the individual abundances as a function of time starting from an initial composition. Rate coefficients for chemical reactions in interstellar medium can be taken from publicly available astrochemical databases such as UdfA \citep{1991A&AS...87..585M, 1997A&AS..121..139M, 2000A&AS..146..157L, 2007A&A...466.1197W} and KIDA \citep{2012ApJS..199...21W, 2015ApJS..217...20W}. The UdfA database, also called the UMIST database, was first released in 1991, followed by subsequent releases in 1995, 1999, and 2006 (http://www.udfa.net/). The KIDA database was first released in 2012 and an updated version was released in 2015 (https://kida.astrochem-tools.org/).

The hydrochemical code KM2 \citep{2015ApJ...808...46M} has chemistry and hydrodynamics modules, allowing us to deal with astrophysical problems requiring coupled evolution. KM2 belongs to the Kinetic Modules suite, developed by the ASIAA CHARMS group.
Its hydrodynamics module is finite-volume conservative and treats the chemical species as passively advected scalars, flowing inside an axisymmetric grid in cylindrical coordinates. The chemistry module solves non-equilibrium chemistry and energy changes due to thermal processes by transferring external ultraviolet radiation, including self-shielding effects on the photodissociation of CO and H$_2$. Other than KM2, hydrodynamic simulations coupled with non-equilibrium chemistry are used in studies such as primordial star formation \citep{2016ApJ...824..119H, 2019MNRAS.490..513S} and photoevaporation of protoplanetary disks \citep{2018ApJ...857...57N}. The computational cost of the chemical module grows with the number of reactions and species considered in the reaction network, and it can be dominant in a hybrid hydrochemical simulation because each cell of the hydrodynamic simulation grid needs to have a separate chemical evolution of its many species undergoing numerous reactions. The matrix describing an extensive reaction network is very sparse, allowing a substantial reduction in runtime. Chemical reaction systems that include reactions with widely different reaction rates are represented by stiff differential equations, and the stiffness of the equations is also an important factor affecting computational cost. The larger the chemical reaction network, the stiffer the system, requiring the use of smaller time steps to obtain a numerically stable solution. These stiff differential equations are solved using the ordinary differential equation (ODE) solver that can solve stiff differential equations based on a backward differentiation formula method, such as LSODE and its variants included in ODEPACK \citep{hindmarsh1983odepack, radhakrishnan1993description}.

In this work, we focus on the efficient computation of the reaction network utilizing the resources of various sparse matrix storage methods and computer settings and experimentally find the more efficient methods. Sparse matrices have only a small fraction of nonzero elements. They can be efficiently stored and allow efficient algorithms. 
Algorithms traversing an efficiently stored sparse matrix benefit from the concentration of nonzero elements, which reduces the operation count by reducing or eliminating the zeros. Additionally, an algorithm able to traverse the sparse matrix in storage order would also benefit from the increase in data locality, an important factor in improving memory access speed. The choice of storage order and the path of the algorithm through the data can, therefore, influence computational efficiency.

We measure these beneficial effects by comparing the efficiency of different storage mechanisms in their natural traversal order to compute the evolution of the chemical system, defined by a very sparse stoichiometric matrix.

The paper is organized as follows. In Section \ref{sec:algorithm}, we describe an algorithm of our new method and sparse matrix storage formats used in numerical experiments in detail. In Section \ref{sec:settings}, we describe setups of the numerical experiments. In Section \ref{sec:results}, we present and analyze our experimental results. In Section \ref{sec:discussion} and \ref{sec:summary}, we discuss our results and summarize our main conclusions.

% #################################################
%  Numerical algorithm
% #################################################

\section{Numerical algorithm}\label{sec:algorithm}

This section describes our new algorithm to solve chemical evolution in interstellar medium.

\subsection{Formulation}
Astrochemical simulations reveal the time evolution of chemical abundance in the interstellar medium.
The production rate of chemical species is written as 
  \begin{eqnarray}
    \frac{{\rm d}n_i}{{\rm d}t} = \mathop{\sum}_{l}\zeta_{li}n_{l} +
     \mathop{\sum}_{j} \mathop{\sum}_{m} k_{jmi} n_{j} n_{m} \nonumber \\ 
      - n_{i}\left (
     \mathop{\sum}_{l} \zeta_{il} + \mathop{\sum}_{m} \mathop{\sum}_{j}
     k_{\rm ijm} n_{j} \right ), 
    \label{eq:modelrate}
  \end{eqnarray}
where $n_i$ is the number density of $i$-th chemical species, $k_{jmi}$ is the reaction coefficient of two body reactions in which $j$-th and $m$-th chemical species react to produce $i$-th chemical species, $\zeta_{li}$ is reaction coefficient of photoreaction or cosmic ray reaction in which $l$-th chemical species produces $i$-th chemical species.

We introduce a matrix formulation to solve simultaneous ordinary differential equations expressed by equation (\ref{eq:modelrate}). 
As an example, let us consider a simple chemical reaction network consisting of the following two reactions:
  \begin{equation}
      2\mathrm{A}   \rightarrow \mathrm{B}, 
  \end{equation}
  \begin{equation}
      \mathrm{B}  + h \nu \rightarrow  2\mathrm{A}.  
  \end{equation}
For simplicity, we denote $k_{AAB}$ as $k$ and $\zeta_{BA}$ as $\zeta$. 
Rates of these reactions are expressed as  $k \, n_A^2$ and $\zeta \, n_B$, respectively. Production rates of each species are given by
\begin{equation}
    \frac{{\rm d}n_{\rm A}}{{\rm d}t} = -2 k \, n_{\rm A}^2 + 2 \zeta \, n_{\rm B}
\end{equation}
\begin{equation}
    \frac{{\rm d}n_{\rm B}}{{\rm d}t} =  k n_{\rm A}^2 - \zeta n_{\rm B} 
\end{equation}
This set of these equations can be equivalently expressed in matrix form:
\begin{eqnarray}
    \frac{\rm d}{{\rm d} t} 
    \left(
    \begin{array}{c}
      n_A \\
      n_B 
    \end{array}     
    \right)
    =
   \left(
   \begin{array}{cccc} 
   -2 & 2  \\ 
   1 &  -1  
   \end{array} \right)
    \left(
    \begin{array}{c}
      k n_{\rm A}^2 \\
      \zeta n_{\rm B} 
    \end{array}
  \right).    
\end{eqnarray}
Generalization of this equation for chemical reaction network with m reactions and n reacting species yields   
\begin{eqnarray}
    \frac{{\rm d}}{{\rm d}t} \left(
    \begin{array}{c}
      n_1 \\
      n_2 \\
      \vdots \\
      n_n
    \end{array}
  \right) 
  =
   \left(
   \begin{array}{cccc} 
   \alpha_{11} & \alpha_{12} &  \cdots & \alpha_{1m} \\
   \alpha_{21} & \alpha_{22} &  \cdots & \alpha_{2m} \\
   \vdots & \vdots & \ddots & \vdots \\
   \alpha_{n1} & \alpha_{n2}  & \cdots & \alpha_{nm}
   \end{array} \right)
    \left(
    \begin{array}{c}
      R_1 \\
      R_2 \\
      \vdots \\
      R_m
    \end{array}
  \right)
    \label{eq:matrix-form}
\end{eqnarray}
where $R_j$ is the reaction rate for $j$-th reaction. The $\alpha_{ij}$ value represents how much $i$-th species increase or decrease in $j$-th reaction. It is negative for reactants, positive for products, and zero for species not involved in the reaction.  The equation's matrix on the right-hand side is called the stoichiometry matrix. 

\subsection{Sparse Matrix Storage Formats}
Matrix-vector multiplication in equation (\ref{eq:matrix-form}) is the most computationally expensive part in simulations with matrix formulation. Efficient sparse matrix-vector multiplication is key for performing astrochemical simulations with our algorithm. The stoichiometry matrix for large reaction networks is generally sparse, {i.e.}, most of the elements have a zero value. 
Various sparse matrix storage formats have been proposed to improve computational efficiency by avoiding operations on zero elements as much as possible, but which one is suitable depends on the number and distribution of nonzero elements in the matrix. In this paper, to find a suitable sparse matrix storage format for astrochemical simulations, we compare the performance of astrochemical simulations with four common storage formats: coordinate (COO) format, compressed column storage (CCS) format, compressed row storage (CRS) format, and Sliced ELLPACK (SELL) format. 

Let us consider following sparse matrix-vector multiplication to illustrate these formats.
\begin{eqnarray}
  \left(
    \begin{array}{c}
      y_1 \\
      y_2 \\
      y_3 \\
      y_4
    \end{array}
  \right) 
  =
   \left(
   \begin{array}{cccc} 
   1 & 0 & 0 & 2 \\
   0 & 0 & 3 & 0 \\
   0 & 4 & 5 & 6 \\
   7 & 0 & 0 & 0
   \end{array} \right)
    \left(
    \begin{array}{c}
      x_1 \\
      x_2 \\
      x_3 \\
      x_4
    \end{array}
  \right)
    \label{eq:example-SpMV}
\end{eqnarray}

Listing \ref{list:coo} shows example Fortran code and data structure for COO format. 
Values of nonzero elements are stored in one dimensional array $A$ in arbitrary order. Row and column indices of nonzero elements are stored in $row\_ind$ and $col\_ind$, respectively. Note that Fortran array index starts at 1, not at 0. Total number of nonzero elements is denoted by $n_{nz}$.

\begin{lstlisting}[language=fortran, caption=Example code segment for COO format in Fortran 90., label=list:coo]
A =(/3,6,1,7,4,5,2/)
col_ind = (/3,4,1,1,2,3,4/)
row_ind = (/2,3,1,4,3,3,1/)
n_nz = 7

y(1:4) = 0
do k = 1, n_nz
    y(row_ind(k)) = y(row_ind(k)) &
   &                 + A(k) * x(col_ind(k))
end do
\end{lstlisting}

\begin{figure}
  \includegraphics[width=\columnwidth]{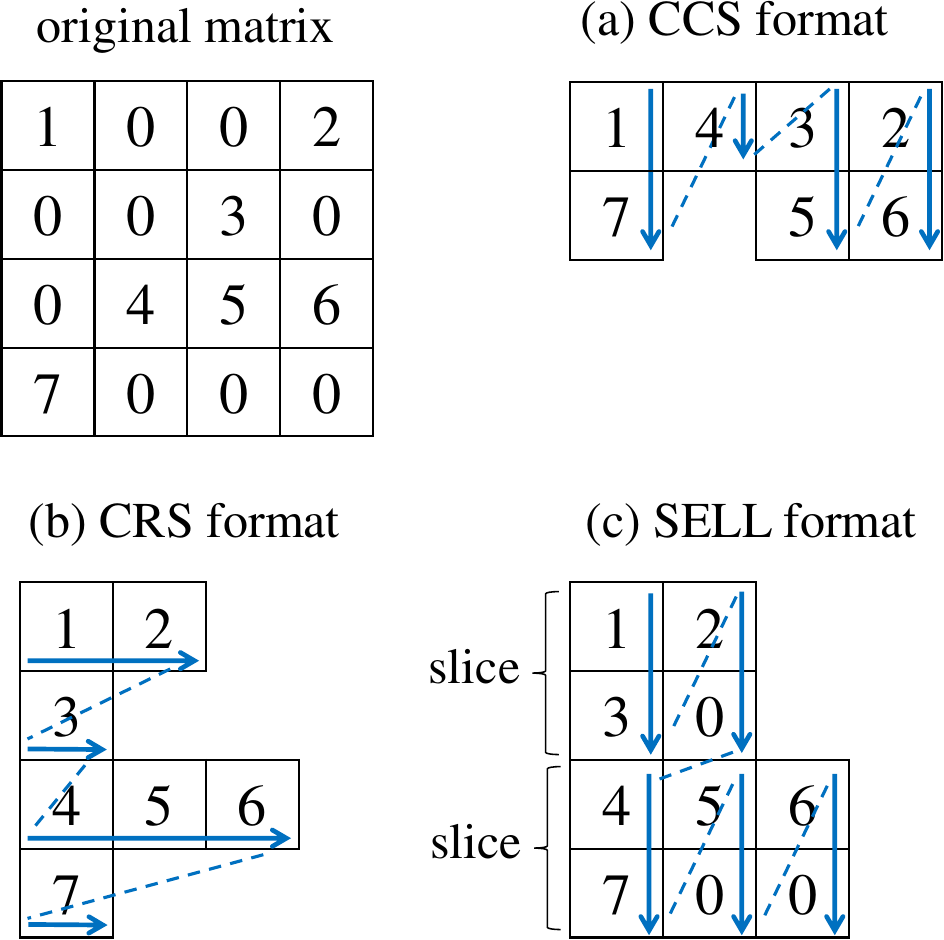}
  \caption{Orders in which nonzero elements are calculated in CCS, CRS, and SELL formats.}
  \label{fig:schematic}
\end{figure}

Fig.\ \ref{fig:schematic}(a) schematically illustrates the order in which nonzero elements are calculated in CCS format, and Listing \ref{list:ccs} shows an example of the Fortran code. 
Values of nonzero elements are stored in one-dimensional array $A$ in column-wise order. Row indices of nonzero elements are stored in $row\_ind$. Index pointers to the first nonzero element of each column in $A$ are stored in $col\_ptr$. Total number of columns is denoted by $n_{col}$.

\begin{lstlisting}[language=fortran, caption=Example Fortran code and data structure for CCS format., label=list:ccs]
A = (/1,7,4,3,5,2,6/)
row_ind = (/1,4,3,2,3,1,3/)
col_ptr = (/1,3,4,6,8/)
n_col = 4

y(1:4) = 0
do j = 1, n_col
  do i = col_ptr(j), col_ptr(j+1) - 1
    y(row_ind(i)) = y(row_ind(i)) + A(i) * x(j)
  end do
end do
\end{lstlisting}

Fig.\ \ref{fig:schematic}(b) schematically illustrates the order in which nonzero elements are calculated in CRS format, and Listing \ref{list:crs} shows an example of Fortran code. 
Values of nonzero elements are stored in an array $A$ and calculated in row-wise order. Column indices of nonzero elements are stored in an array $col\_ind$. Index pointers to the first nonzero element of each column in $A$ are stored in an array  $row\_ptr$. Total number of rows is denoted by $n_{row}$.

\begin{lstlisting}[language=fortran, caption=Example Fortran code and data structure for CRS format., label=list:crs]
A = (/1,2,3,4,5,6,7/)
col_ind = (/1,4,3,2,3,4,1/)
row_ptr = (/1,3,4,7,8/)
n_row = 4

y(1:4) = 0
do i = 1, n_row
  do j = row_ptr(i), row_ptr(i+1) - 1
    y(i) = y(i) + A(j) * x(col_ind(j))
  end do
end do
\end{lstlisting}

Fig.\ \ref{fig:schematic}(c) schematically illustrates the order in which nonzero elements are calculated in SELL format  \citep{monakov2010}, and Listing \ref{list:sell} shows example Fortran code.
Nonzero elements are shifted to the left side in the matrix and divided into slices of width $wid\_sl$ in the column direction. 
The row length of a slice is the length of row with the largest number of nonzero elements in that slice. 
The row with fewer nonzero elements than that is padded with zero for a uniform row length in the slice. 
Lengths of slices are stored in an array $len\_sl$.
In each slice, values of nonzero elements are stored in an array $A$ in column-wise order.
Column indices of nonzero elements are stored in an array $col\_ind$.
The total number of slices is denoted by $n\_sl$.

\begin{lstlisting}[language=fortran, caption=Example Fortran code and data structure for SELL format., label=list:sell]
A = (/1,3,2,0,4,7,5,0,6,0/)
col_ind = (/1,3,4,4,2,1,3,2,4,3/)
len_sl = (/2,3/)
wid_sl = 2
n_sl = 2

y(1:4) = 0
j_ptr = 1
do i = 1, n_sl
  i_sl = (i-1) * wid_sl
  do j = 1, len_sl(i)
    do k = 1, wid_sl
      y(i_sl + k) = y(i_sl + k) + A(j_ptr) &
     &                  * x(col_ind(j_ptr))
      j_ptr = j_ptr + 1
    end do
  end do
end do
\end{lstlisting}

\begin{figure}
	\includegraphics[width=\columnwidth]{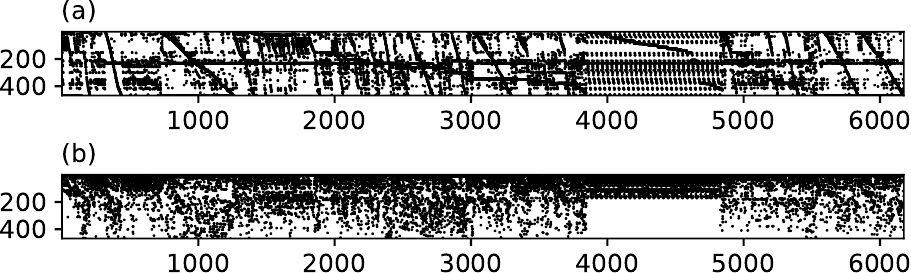}
    \caption{Distributions of nonzero elements in stoichiometry matrix without row reordering (a) and with row reordering (b).}
    \label{fig:distribution}
\end{figure}

% #################################################
%  Experimental Settings
% #################################################

\section{Experimental Settings}\label{sec:settings}
In this section, we describe the experimental settings for performance comparison study of different sparse matrix storage formats. 
We implemented our proposed algorithm in hydrochemical hybrid code KM2 \citep{2015ApJ...808...46M} and simulated chemical evolution in photon dominated region (PDR) as a benchmark test.  
The hydro module of KM2 was temporarily turned off for this study. 
Physical parameters for the simulation were adopted from model F2 in \cite{2007A&A...467..187R}. 
A semi-infinite plane parallel cloud with the number density of $n_{\mathrm{H}} = 10^3$ $\mathrm{cm^{-3}}$ was illuminated by FUV radiation with intensity of $\chi=10^5$ in units of the Draine field. 
The gas and dust temperatures were fixed at 50 K and 20 K, respectively. 
Visual extinction $A_V$ was assumed to be related to  hydrogen column density $N_{\mathrm{H},tot}$ by $A_V = 6.289 \times 10^{-22} N_{\mathrm{H},tot}$.
All elements in the gas were atomic state under the initial condition. 
Elemental abundances adopted in this study are summarized in Table \ref{tab:abundance}. 
For the elements included in the model F2 in \cite{2007A&A...467..187R}, i.e. H, He, C, and O, the same abundances as in \cite{2007A&A...467..187R} were used, abundances of other elements were taken from \cite{2008ApJ...682..283G}. 
Chemical evolution solved for 30 Myr. 
A logarithmic grid was used to obtain high resolution near the cloud surface, where chemical reactions are strongly affected by FUV. 
Computational domain covers the range of $0 < A_V < 20$  with 128 cells. 
For further details on benchmark, refer to \cite{2007A&A...467..187R} and \cite{2015ApJ...808...46M}.

The chemical reaction network of this performance comparison experiment consists of 6175 reactions among 468 species involving 13 elements taken from UMIST rate12 database \citep{2013A&A...550A..36M}.
The stoichiometry matrix of the chemical reaction network is 468 $\times$ 6175  matrix. 
Total number of nonzero elements is 24580, which is about 0.085 \% of the total number of matrix elements. 
Each row corresponds to a chemical species, and each column corresponds to a reaction. 
The order of rows and columns can be arbitrary. 
Two stoichiometry matrices with different row order were used to see how distribution of nonzero elements affects computing 
performance. 
One stoichiometry matrix is created by using the same order of chemical species and chemical reactions in UMIST rate12 database, that is,  
chemical species are in alphabetical order and chemical reactions are grouped by reaction types such as charge exchange, dissociative recombination, photoprocess (hereinafter referred to as original matrix). 
The other stoichiometry matrix is created by reordering the rows of the above matrix in order of the number of nonzero elements (hereinafter referred to as reordered matrix). 
Fig \ref{fig:distribution} (a) and (b) show distributions of nonzero elements in the original matrix and the reordered matrix, respectively. 

\begin{table}
   \centering
   \caption{Elemental abundances relative to hydrogen nuclei}
   \label{tab:abundance}
   \begin{tabular}{lc|lc}
        \hline \hline
           Element &  Abundance &  Element & Abundance \\ \hline
           H  &  1.0 &  Si &  $8.0 \times 10^{-9}$ \\ 
           He &  0.1 & Mg & $7.0 \times 10^{-9}$  \\ 
           O  &  $3.0 \times 10^{-4}$ & Cl & $4.0 \times 10^{-9}$ \\ 
           C  &  $1.0 \times 10^{-4}$ & Fe & $3.0 \times 10^{-9}$ \\ 
           N  &  $7.5 \times 10^{-5}$ & P  & $3.0 \times 10^{-9}$ \\
           S  &  $8.0 \times 10^{-8}$ & Na & $2.0 \times 10^{-9}$ \\
           F  &  $2.0 \times 10^{-8}$ &  &  \\
       \hline
     \end{tabular}
 \end{table}
 
Recent CPUs achieve high performance by incorporating Single Instruction Multiple Data (SIMD) units that can perform the same operation on multiple data concurrently. In addition to the effect of the distribution of non-zero elements, simulations using each sparse matrix storage format were performed with three different settings to investigate the effect of SIMD operations. 
In the first setting, the reordered matrix was used and SIMD operation of Advanced Vector Extensions 2 (AVX2) was enabled with compilation option -O3 -xcore-avx2. 
AVX2 instructions enable CPUs to operate four double-precision floating-point numbers concurrently.
In the second setting, the original matrix was used and SIMD operation was enabled with compilation option -O3 -xcore-avx2. 
In the third setting, the reordered matrix was used and SIMD operation was partly disabled with compilation option -no-simd -no-vec in the critical routines. 
These simulations were run in serial. 

Specification of computing node that we used is summarized in Tale \ref{tab:environment}.
Our computing node has dual CPU and each CPU has 10 physical cores. 
We measured not only execution times of the simulations in serial but also execution times of the simulations in parallel using OpenMP. 
Parallelization using OpenMP was done with different levels of task granularity. 
One is parallelization at the loop that computes production rates of chemical species as shown in Listings \ref{list:coo} to \ref{list:sell}. 
The other is a more coarse-grained parallelization by domain decomposition approach, that is, the entire computational domain is divided into smaller subdomains. 
In these parallelized simulations, reordered matrix was used and SIMD operations were enabled. 
Thread affinity policy was specified by setting environment variables as OMP\_PLACES=cores and OMP\_PROC\_BIND=close. 
Considering the influence of background processes, simulations were run five times each, and averages and standard errors of their execution times were obtained.

\begin{table}
   \centering
   \caption{Specification of computing node}
   \label{tab:environment}
   \begin{tabular}{|l|l|}
        \hline
           CPU & Intel Xeon E5-2650 v3 2.3 GHz $\times$ 2\\
           &  Intel Haswell microarchitecture \\ \hline
       Memory  &  DDR4-2133 Dual-Channel 32GB  \\ \hline
       OS      &  Ubuntu 16 \\ \hline
        Compiler    &  Intel Fortran compiler 16.0 \\
       \hline
     \end{tabular}
 \end{table}
 
\begin{table}
   \centering
   \caption{Specification of Kawas computing nodes}
   \label{tab:kawas}
   \begin{tabular}{|l|l|}
        \hline
           CPU & AMD EPYC 7763 @ 2.45GHz \\
           & 64-Core Processor $\times$ 2 (128 Cores) \\ \hline
       Memory  &  512GB \\ \hline
       OS      &  Rocky Linux x86\_64 8.5 \\ \hline
        Compiler    &  Intel Fortran compiler 22.1 \\
       \hline
     \end{tabular}
 \end{table}

As a reference, simulations with conventional algorithm were also run by using KROME package \citep{2014MNRAS.439.2386G}. 
In the KROME package, there are two options to compute the right-hand side of equation (\ref{eq:modelrate}). One is the explicit method and the other is the implicit method. We used the explicit method, which performs better. KM2 and KROME use the same stiff ODE solver, DLSODES, which is a double-precision version of LSODES, a variant of basic solver LSODE in ODEPACK. DLSODES is intended for problems in which the Jacobian matrix is sparse. Relative and absolute tolerances to control the magnitude of acceptable errors were set to $1.0 \times 10^{-4}$ and $1.0 \times 10^{-20}$ for both of KM2 and KROME, respectively. The reaction coefficients in equation (\ref{eq:modelrate}) depend on local physical properties such as the temperature, cosmic ray ionization rate or FUV intensity. In the KM2 code, these reaction coefficients were precomputed outside the stiff ODE solver. There is a KROME option `\textit{-useCustomCoe}' for computing reaction coefficients with a user-defined function. By calling the function outside the stiff ODE solver, a similar computational approach to the KM2 can be achieved. It is faster for this purpose than the default KROME settings. 
In conventional algorithms, computations of production rates of chemical species are not expressed by loop iteration and cannot be parallelized at the chemical loop level by OpenMP.

% #################################################
%  Results
% #################################################

\section{Results}\label{sec:results}
\begin{figure}
	\includegraphics[width=\columnwidth]{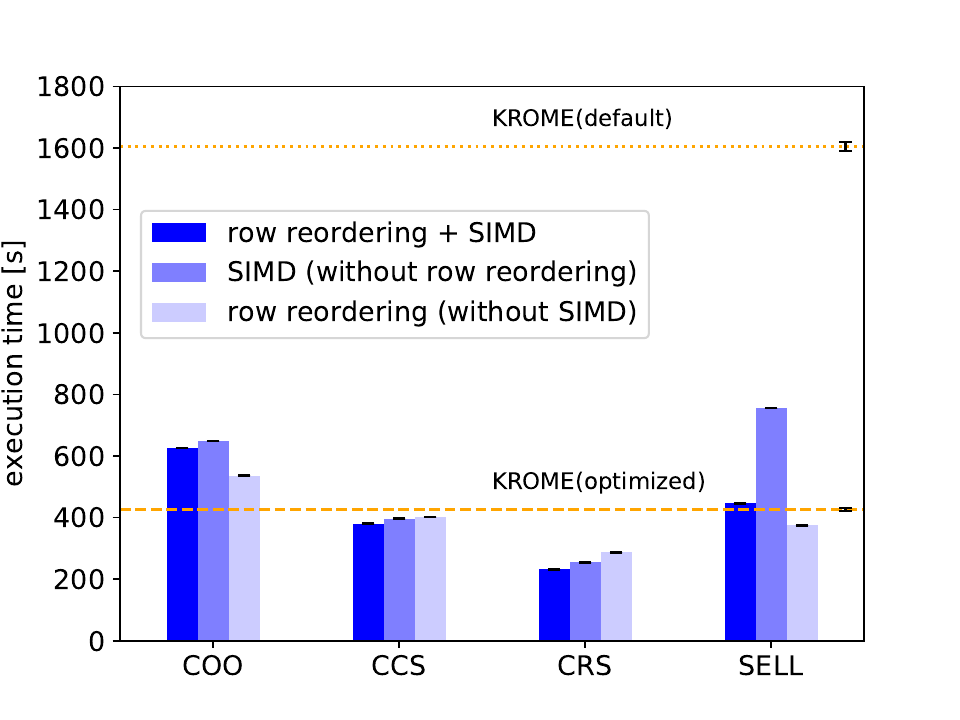}
    \caption{Execution times of simulations with COO, CCS, CRS, and SELL formats in serial. Error bars represent the standard errors of execution times. The dashed and dotted lines represent execution times of reference model using KROME code with SIMD operations and \textit{-useCustomCoe} option, and simulation using the default KROME settings, respectively}.
    \label{fig:serial_results}
\end{figure}

In this section, we describe results of performance comparison experiments.
Fig.\ \ref{fig:serial_results} shows execution times of simulations in serial. 
The execution time of the reference model using KROME with SIMD operations and \textit{-useCustomCoe} option was 427$\,\mathrm{s}$, much faster than default KROME at $1604\,\mathrm{s}$. All simulations with our algorithm showed better performances than the default KROME, and formats CRS (and CCS) were also faster than the reference. 
The CRS format with row reordering and SIMD operations showed the best performance. 
Its execution time was 232 $\mathrm{s}$, which is 6.91 times faster than default KROME and 1.84 times faster than the reference. 
In all settings, CRS format showed the better computing performances than other formats.

Row reordering improved computational performance by more than 10 percent for CRS and SELL sparse matrix storage formats. 
The execution times with row reordering were shorter than those without it. 
The performance improvement rates calculated by dividing the execution time without row reordering by the execution time with row reordering for COO, CCS, CRS, and SELL formats were 1.04, 1.04, 1.10, and 1.69, respectively. 
The performance improvement was especially noticeable for the SELL format.

SIMD operation was effective for CRS format.
The simulation with SIMD operations was faster than that without it by 1.24 times. 
On the other hand, simulation execution time with SIMD operations for CCS format was almost the same as that without it. 
The performance improvement rate was only 1.05. 
SIMD operations caused performance degradation for COO and SELL formats, with the performance improvement rates of 0.86 and 0.84, respectively.
The effects of SIMD operations were also investigated for the reference model.
The simulation execution time without SIMD operations was 1590 $\mathrm{s}$.
The performance improvement rate due to SIMD operations was only 1.01, and the difference was within standard error of measured simulation execution times.

\begin{figure}
	
    \includegraphics[width=\columnwidth]{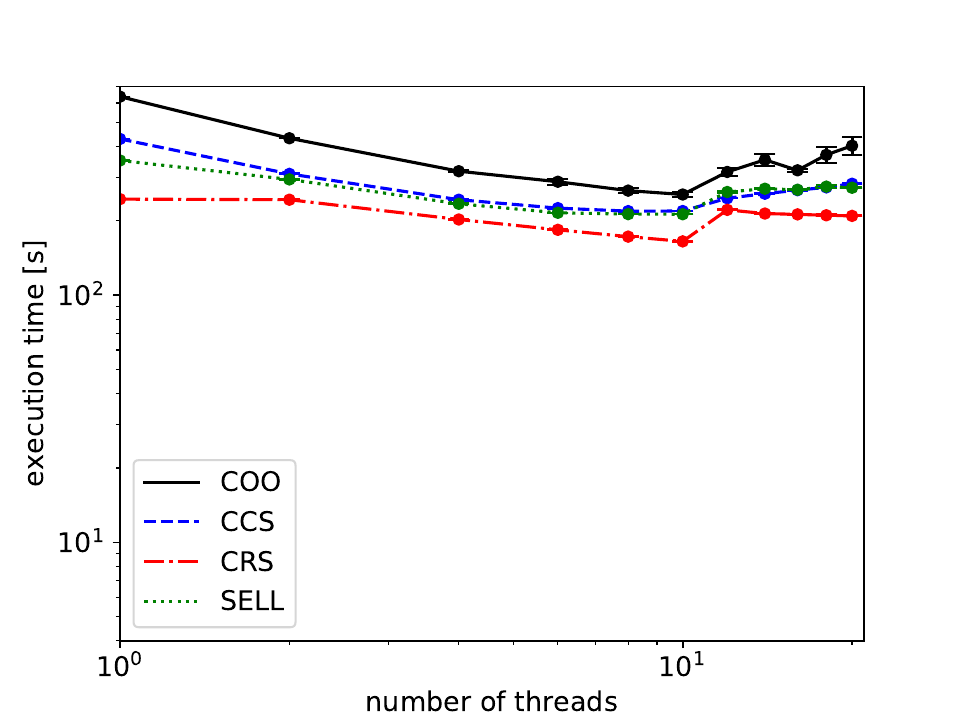}
    \caption{Execution times of simulations as a function of OpenMP threads, for a chemical-loop parallelization. The solid, dashed, dash-dotted, and dotted lines represent COO, CCS, CRS, and SELL formats, respectively. Error bars represent standard errors.}
    \label{fig:openmp_results}
\end{figure}

Fig.\ \ref{fig:openmp_results} shows execution times of simulations parallelized at the loop that computes production rates of chemical species as a functions of number of threads. 
The trend of speed-up by parallel computation differed depending on whether the number of threads exceeded 10, the number of cores in the CPU. 
The execution time of all formats reduced monotonically up to 10 threads within the margin of error.  
On the other hand, the execution times of all formats increased when the number of threads exceeded 10, and almost no speed-up was observed for a higher number of threads. 
At the numbers of threads with the highest performance, the speed-up for COO, CRS, CCS, and SELL formats compared to simulations in serial are 2.34, 1.41, 1.81, and 2.10, respectively. 
The simulation with the CRS format showed the best performance among all formats. 

\begin{figure}
	
    \includegraphics[width=\columnwidth]{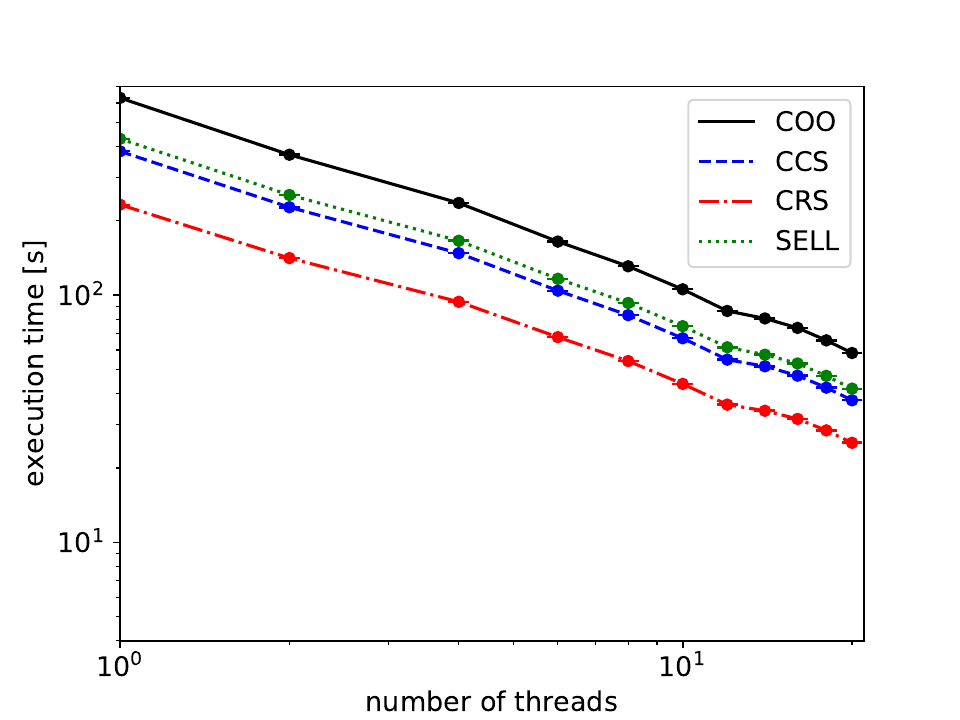}
    \caption{Execution times of simulations parallelized by domain decomposition as a function of OpenMP threads. The solid, dashed, dash-dotted, and dotted lines represent COO, CCS, CRS, and SELL formats, respectively.
    %Error bars represent standard errors.
    }
    \label{fig:domain-decomposition}
\end{figure}

Fig.\ \ref{fig:domain-decomposition} shows execution times of simulations parallelized by domain decomposition as a function of OpenMP threads.
Unlike parallelization at the chemical loop level, execution times reduced monotonically even when number of threads exceeds 10 and the threads spanned multiple CPUs. 
The speed-up for COO, CRS, CCS, and SELL formats compared to simulations in serial are 10.7, 9.16, 10.1, and 10.7, respectively.

In addition to execution times, compilation times were also measured. 
The average compilation time of the default KROME code was 6866 
seconds with a standard error of 19.2 seconds. 
The compilation time of our code was almost independent of the compilation options. 
The mean compilation time of our code for all compilation options was 53.6 seconds with a standard error of 0.4 seconds.

\begin{figure}
    
    \includegraphics[width=\columnwidth]{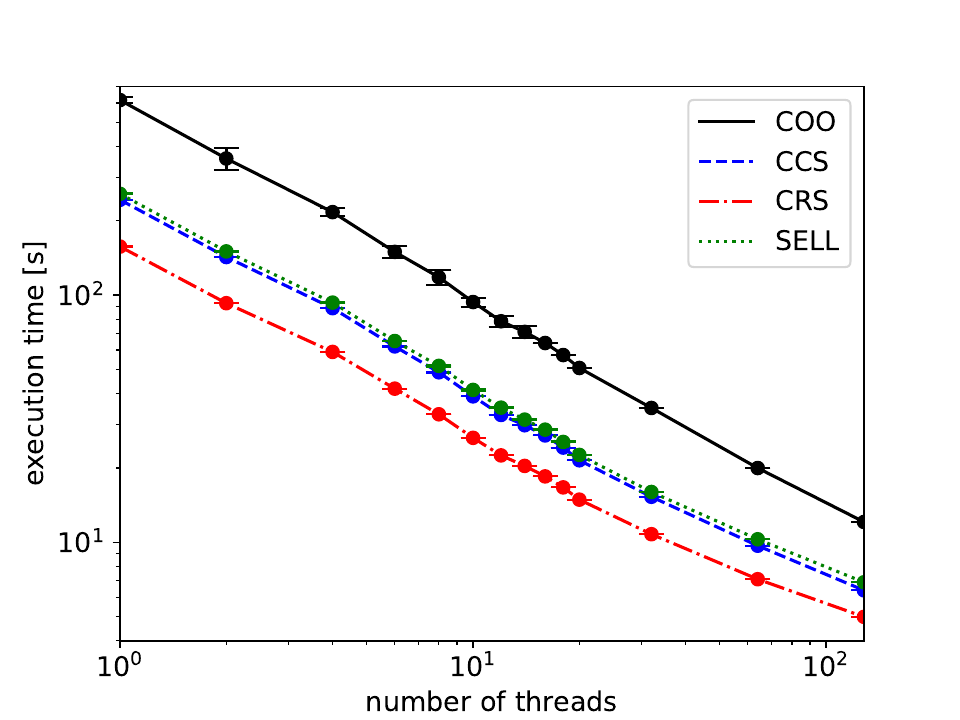}
    \caption{Execution times of simulations parallelized by domain decomposition as a function of OpenMP threads. The solid, dashed, dash-dotted, and dotted lines represent COO, CCS, CRS, and SELL formats, respectively. Simulations performed on the machine Kawas able to efficiently run 128 threads on each of its nodes.
    %Linear and log plots are shown.
    }
    \label{fig:kawas-domain-decomposition}
\end{figure}

We then performed domain-decomposition OpenMP experiments on the large nodes of the machine Kawas available to us at ASIAA\@. Each node of this machine is described on Table \ref{tab:kawas}. We have run tests similar to those of Fig.\ \ref{fig:domain-decomposition} but now also including 32, 64, and 128 threads. Fig.\ \ref{fig:kawas-domain-decomposition} shows the results, which confirm efficiency of domain decomposition up to large number of threads. They also confirm the higher efficiency of the CRS algorithm over the other storage methods.

% #################################################
%  Discussion
% #################################################

\section{Discussion}\label{sec:discussion}

\begin{figure}
	
    \includegraphics[width=\columnwidth]{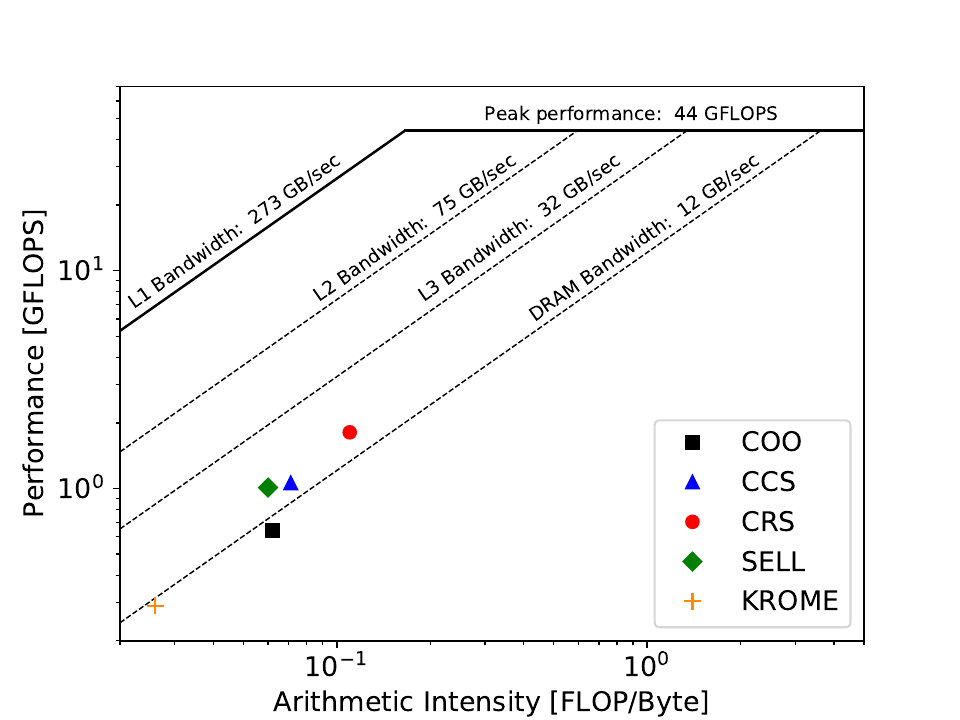}
    \caption{The computational performances of KM2 simulations using reordered matrix with SIMD operations enabled are plotted as function of arithmetic intensity. Additionally, the plus symbol represents computational performance of reference model using KROME with \textit{-useCustomCoe} option. The horizontal line represents peak performance of double-precision floating-point operations, and the diagonal lines represent bandwidth of memory and the first-level, second-level, third-level caches on our computing node.
    }
    \label{fig:roofline}
\end{figure}

The results of the performance comparison study show that our new algorithm significantly improves the performance of astrochemical simulation compared to the conventional algorithm. 
The new algorithm using CRS format is particularly suited for astrochemical simulations. 
It also has the advantage of greatly reducing compilation time. 
Our algorithm is easy to implement and can improve the performance of existing simulation codes. 
While performance was greatly improved for simulations in serial, and also in OpenMP-parallel by applying domain decomposition, high parallelization efficiency was not achieved for simulations in parallel with OpenMP applied to the chemical loops. 
Roofline performance model \citep{2009ACM...52...65W} is useful to see which hardware component limits application performance. Fig. \ref{fig:roofline} shows the results of roofline analysis obtained by using Intel\textregistered Advisor for the simulations in serial. Computational performances are plotted as function of arithmetic intensity, which is the ratio of total floating-point operations to the number of bytes accessed in the main memory. On our computing node, applications with arithmetic intensity greater than 0.166 FLOP Byte$^{-1}$ can be considered as CPU bound, while those below this threshold can be considered as memory bound. All plotted points are within the memory-bound area, indicating that astrochemical simulations are memory-bound regardless of whether the conventional or our new algorithm are employed. Effective utilization of cache and increased arithmetic intensity are important factors for high computational performance. The performances of the reference model using KROME with \textit{-useCustomCoe} option and the simulation with COO format are lower than the limitation imposed by memory bandwidth. These simulations hardly take advantage of the cache. On the other hand, simulations with CRS, CCS, and SELL format exhibit performance higher than this limitation, leveraging the advantage of cache to some extent. 
We conducted performance profiling by Intel\textregistered VTune for simulations employing reordered matrix with SIMD operations enabled. The computational time taken by stiff ODE solver exceeded 97.9 percent of the total computation time for simulations using any sparse matrix format. Sparse matrix operations constitute the most time consuming part within the stiff ODE solver, accounting for fractions of total computation time of 85.3\%, 75.7\%, 60.5\%, and 79.1\% with COO, CCS, CRS, and SELL format, respectively. CRS format shows the best performance for sparse matrix operations. Profiling clearly shows the steep relative reduction in computation time fraction for the best of these four formats.

The differences in computation time by reordering the rows of the stoichiometry matrix indicate that the performance of our algorithm depends on the distribution of nonzero elements.
This is to be expected since amount of random memory access due to irregular distribution of nonzero elements is an important factor for performance of sparse matrix-vector multiplication.
Reordering the rows in order of the number of nonzero elements increases the locality of the nonzero elements. 
This improves cache hit ratio because the data used in computation is more likely to line up in contiguous locations in memory. 
In addition, in the SELL format, zero padding is reduced by having rows with a close number of nonzero elements in the slice, resulting in a degree of performance improvement.

The degree of simulation speed-up by SIMD operations depends on the sparse matrix storage format, which can be explained by the characteristic of the matrix used in the simulations and effects of indirect memory reference. 
As described in Section \ref{sec:settings}, stoichiometry matrices used in astrochemical simulation are much longer in the row direction than in the column direction, and therefore have more non-zero elements. 
Atomic hydrogen is involved in 1794 reactions, and the row representing atomic hydrogen has the highest number of nonzero elements. 
Next to this, molecular hydrogen is involved in 980 reactions and electron in 891 reactions. 
Since CRS format stores nonzero elements in row-wise order, simulations with CRS format have longer loop length for the innermost loop than simulation with other storage formats. 
Thanks to the long loop length, the speed-up of SIMD operations is superior to the overhead of SIMD operations.
Simulations using the CCS or SELL format have short loop lengths, which results in a large overhead relative to the speed-up effect of SIMD operations.
The innermost loop length in simulations with CCS format is the number of chemical species in reaction, which is at most 6 in UMIST rate12 database. 
The innermost loop length in simulations with SELL format is width of slice, which is 4 in this experiment. 
In simulations with format other than the COO format, there is only one indirect memory reference in the loop, but in simulations with the COO format, there are two, i.e., indices of column and row. 
Indirect memory references in loops make it difficult to achieve the speed-up effect of SIMD operations. 
It is probably because the two indirect memory references have a larger overhead than the speed-up by SIMD operations, and the simulation with the COO format degrades performance when SIMD operations are enabled. 
In the model using the KROME code (with default options), the simulation is not sped up at all even with SIMD operations enabled.

Simulations parallelized at the chemical loop showed performance degradation for larger numbers of threads (larger than 10 in our experiment due to machine architecture).
This is because the chemical rate computation is memory-bound, not CPU-bound.  
On the other hand, the domain-decomposition parallelization achieves good speed-ups essentially for all numbers of threads. This is because (1) a larger fraction of the problem is parallelized and not only the chemical rate calculation, and (2) there are fewer issues of contention for the usage of memory because each thread is now operating on largely independent sets of data. We have shown that this hybrid code can be parallelized at both levels, chemical loop and domain decomposition, each one to be preferred depending on the size of the domain and number of threads.
The study of this domain-decomposition parallelization, natural to KM2, serves also as an example of the application of sparse-matrix techniques to a chemical code that belongs to an already-established parallel framework and thus is not in immediate need of a more internal parallelization.

%\Add{During the development of the two parallelization models (domain-decomposition and chemical-loop based) we timed different portions of the code. This was done primarily to guide the efficient parallelization. As a side-effect we obtained information about code timing of certain portions of the code. We measured that the code spends most of its time in the main loop. The dpmain-decomposition level parallelization applies to nearly the totality of the running time. The chemical-loop parallelization was measured to speed-up routines taking up about $2/3$ of the running time, which is a considerable fraction but still short of the totality. This is because some parallelizable routines of the main loop are outside the chemical-loop parallelization.}

The temperature was assumed to be constant in our performance comparison experiments, which is a reasonable approximation for star formation processes, one of the key targets of astrochemical simulations. As described in \cite{2015ApJ...808...46M}, the chemistry module of the KM2 determines the time step so that the temperature change from the previous step is small enough, and updates the chemical abundances separately from the temperature in the usual operator-split manner. The results of our performance comparison experiments are directly applicable to cases of variable temperature. For other chemistry codes that solve the temperature and chemical abundances together within a stiff ODE solver, the computational cost of updating the chemical abundances is significantly higher than that of updating the temperature. Therefore a large performance improvement is expected by using our new algorithm as well.

A more general conclusion of this work is that sparse-matrix techniques are efficient and effective. They are not extremely difficult to implement, although their most efficient usage on a given computer architecture often needs attention to memory distribution and CPU usage.
We have observed that sparse-matrix methods are already widely used in many fields of computer applications. However, they are still under-utilized compared to their full potential applications. That may be because the use of full or nearly full matrices instead of sparse matrices can also be effective even if not the most efficient method, and it can be easier to implement.
Within astrophysics, we observe that sparse-matrix methods can be applied to astrochemistry, to models of grain coagulation, and to problems in nucleosynthesis, without exhausting the list of potential applications. Matrices describing models are often sparse, a fact not always completely exploited for computational advantage.

% #################################################
%  Summary
% #################################################

\section{Summary}\label{sec:summary} % or maybe have a "discussion and summary" section instead
We advance a substantial speed-up of astrochemical computations by making use of their natural sparse matrix structure.
Up to 1.84$\times$ speed-up was found when compared to a conventional approach (KROME with \textit{-useCustomCoe} option), and up to 6.9$\times$ when compared with the default KROME options.
Several sparse-matrix algorithm variations were tried, showing the advantages of (1) reordering the matrix to improve locality, (2) using sparse-matrix storage that compresses along the less sparse row direction of this matrix, (3) using SIMD operations. OpenMP parallelization was explored with two techniques (chemical parallelization and domain decomposition), each having its advantages
for problems with different sizes of chemistry or geometry.

\begin{acknowledgments}
This work is supported by the Astrobiology Center Program of
National Institutes of Natural Sciences (NINS) Grant Number AB271010, JSPS KAKENHI Grant Number JP23K03457, and a University Research Support Grant from the National Astronomical Observatory of Japan (NAOJ). Numerical analyses were in part carried out on computers at the Center for Computational Astrophysics, National Astronomical Observatory of Japan. We acknowledge previous support by the National Institute of Informatics, Japan.
The authors acknowledge grant support for the CHARMS group from the Institute of Astronomy and Astrophysics, Academia Sinica (ASIAA), and the National Science and Technology Council (NSTC) in Taiwan through grants 111-2112-M-001-074- and 112-2112-M-001-030-. The authors acknowledge the access to high-performance facilities (TIARA cluster and storage) in ASIAA and thank the National Center for High-performance Computing (NCHC) of National Applied Research Laboratories (NARLabs) in Taiwan for providing computational and storage resources. This work utilized tools (Kinetic Modules suite of codes) developed and maintained by the ASIAA CHARMS group. This research has made use of SAO/NASA Astrophysics Data System.
\end{acknowledgments}

\bibliography{references}{}
\bibliographystyle{aasjournal}


\end{document}